\documentclass[amsmath,onecolumn,amssymb,showkeys,superscriptaddress,aps,prd,10pt,onecolumn]{revtex4-2}
\usepackage{graphicx}
\usepackage[caption=false]{subfig}
\usepackage{multirow}
\usepackage{appendix}
\usepackage{graphicx,epstopdf}
\usepackage{subfig}
\usepackage{colortbl}
\usepackage{xcolor}
\usepackage[colorlinks=true, urlcolor=blue, linkcolor=blue, citecolor=blue]{hyperref}
\usepackage{float}
\usepackage{slashed}
\usepackage[utf8x]{inputenc}

\begin{document}

\title{Spectrum of $cc\bar{b}\bar{b}$, $bc\bar{c}\bar{c}$, and $bc\bar{b}\bar{b}$  Tetraquark States in the Dynamical Diquark Model}

\author{Halil Mutuk}
 \email{hmutuk@omu.edu.tr}
 \affiliation{Department of Physics, Faculty of Sciences, Ondokuz Mayis University, 55139 Samsun, Türkiye}
 \affiliation{Institute for Advanced Simulation, Institut für Kernphysik and Jülich Center for Hadron Physics, Forschungszentrum Jülich, D-52425 Jülich, Germany}

\begin{abstract}
The dynamical diquark model assumes that exotic hadrons can be formed from colored diquarks. This model asserts a multiquark exotic state composed of a compact diquark $\delta$ and antidiquark $\bar{\delta}$ for a tetraquark interacting through a gluonic field of finite extent. Using  Born-Oppenheimer (BO) approximation and BO potential calculated numerically on the lattice, we study mass spectra of $S-$wave $cc\bar{b}\bar{b}$, $bc\bar{c}\bar{c}$, and $bc\bar{b}\bar{b}$ tetraquark systems in the basis of diquark spins. We assume that colour-antitriple diquark and colour-triplet antidiquark form tetraquark state. The predicted mass spectrum for ground state of $cc\bar{b}\bar{b}$ is found to be lower then their corresponding two-meson thresholds. This system may be a candidate for bound state. Masses of $bc\bar{c}\bar{c}$ and $bc\bar{b}\bar{b}$ tetraquark states are found to be above than the corresponding two-meson thresholds.
\end{abstract}

%\pacs{Later}% PACS, the Physics and Astronomy
\maketitle
\section{Introduction}
Quantum Chromodynamics (QCD) successfully describes the interactions of quarks and gluons and presents a much richer spectrum of hadrons than the simple baryons and mesons of the quark model. The hadrons that exist outside of the framework of quark model are called exotic hadrons. Searching for exotic hadrons is an interesting research field which is full of challenges and also opportunities. Indeed, theoretical and experimental studies of exotic hadrons beyond the conventional quark model are important test of nonperturbative aspects of  QCD, i.e. low energy regime of QCD. 

Since 2003, many exotic particles have been observed in the experiments \cite{Choi:2007wga,Belle:2011aa,Adachi:2012cx,Chilikin:2013tch,Ablikim:2013xfr,Ablikim:2013wzq,Ablikim:2013mio,Aaij:2014jqa,Chilikin:2014bkk,Aaij:2015tga,Aaij:2018bla} which cannot be fitted in the conventional quark model. There are many possible interpretations to understand the inner structures and physical properties of these exotic particles: molecular picture (loosely bound state of two hadrons), tetraquarks (compact bound states), hybrids (composed of gluons and quarks), hadroquarkonia (composed of a heavy quarkonium embedded in a light meson), etc. For recent reviews of these models, see Refs. \cite{Lebed:2016hpi,Chen:2016qju,Esposito:2016noz,Guo:2017jvc,Ali:2017jda,Liu:2019zoy}. 

In 2017, the CMS Collaboration observed $\Upsilon(1S)$ pair production and indicated a $bb\bar{b}\bar{b}$ signal around 18.4 GeV with a global significance of $3\sigma$ \cite{CMS:2016liw}. The measurement of this signal inspired lots of theoretical studies on the $bb\bar{b}\bar{b}$ tetraquarks \cite{Chen:2016jxd,Esposito:2018cwh,Hughes:2017xie,Karliner:2016zzc,Richard:2017vry,Bai:2016int,Chen:2019dvd}. Later the LHCb Collaboration investigated the $\Upsilon \mu^+ \mu^-$ invariant mass distribution for a possible fully-heavy tetraquark state $bb\bar{b}\bar{b}$ and did not observe the relevant state \cite{Aaij:2018zrb}. Very recently the LHCb Collaboration announced the evidence for new resonance structures; a narrow structure $X(6900)$ around 6.9 GeV in addition to a broad structure in the range of $6.2-6.8$ GeV in di-$J/\psi$ mass spectrum at center of mass energies $\sqrt{s}=7,8$ and 13 TeV \cite{Aaij:2020fnh}. The narrow structure is assumed as a resonance with the Breit-Wigner line shape. The general view for $X(6900)$ structure is to be composed of four charm quarks in the form of $cc\bar{c}\bar{c}$. The observation of these new structures triggered many studies in the literature \cite{Albuquerque:2020hio,An:2020jix,Chen:2019dvd,Chen:2020xwe,Debastiani:2017msn,Feng:2020riv,Giron:2020wpx,Gordillo:2020sgc,Guo:2020pvt,Huang:2021vtb,Jin:2020jfc,Karliner:2020dta,Ke:2021iyh,Li:2019uch,Li:2021ygk,Liang:2021fzr,liu:2020eha,Lu:2020cns,Ma:2020kwb,Pal:2021gkr,Sonnenschein:2020nwn,Wan:2020fsk,Wang:2018poa,Wang:2020gmd,Wang:2020ols,Wang:2020tpt,Wang:2021kfv,Weng:2020jao,Yang:2020rih,Yang:2020wkh,Zhang:2020xtb,Zhao:2020cfi,Zhao:2020nwy,Zhu:2020snb,Zhu:2020xni,Gong:2020bmg,Cao:2020gul,Yang:2021zrc,Mutuk:2021hmi}. 

Among these interpretations for four-quark exotics, tetraquark picture is a topic of great interest, especially when fully-heavy tetraquark $(QQ\bar{Q}\bar{Q})$ is on the scene. They should be predominantly compact tetraquarks rather than molecule since only heavy $(Q\bar{Q})$ mesons can be exchanged between two subsystems in such a molecule. The potential arising between this exchange is the type of Yukawa potential and it is not strong enough for binding. Soft gluons with mass around 0.5 GeV can be exchanged between two-heavy quarkonia yielding QCD van der Waals force. Although this force is to be known as attractive, it is unclear that QCD van der Waals force is strong enough to form a bound state. The hadroquarkonium picture is not applicable since it includes light meson. Therefore the tetraquark picture is favorable for fully-heavy four-quark structures. 

On the other side, it is natural to expect the lowest-lying $Q_1 \bar{Q}_2 Q_3 \bar{Q}_4$ states to have comparable distances between all four heavy quarks. For example, if  the $(Q_1 \bar{Q}_2)$ and $(Q_3 \bar{Q}_4)$ pairs are formed with much smaller internal separations than the distance between the two pairs, then formation of two free traditional quarkonium states occurs rather than a single resonance, even though these $(Q_1 \bar{Q}_2)$ and $(Q_3 \bar{Q}_4)$ pairs are in color octets and need gluon exchange in order to hadronize as color singlets. Indeed, closer association of $Q\bar{Q}$ pairs can yield an immediate dissociation into quarkonium pairs. In this case, $(Q_1 Q_3) (\bar{Q}_2 \bar{Q}_4)$ configuration with color-triplet diquarks is the only one that survives via an attractive interaction between the component constituents, i.e., quarks inside the diquarks, and remain bound due to the confinement with exchange of any number of gluons. This mechanism is quite different than the one-gluon-exchange (OGE) mechanism and these features define the dynamical diquark picture of the multiquark exotics. The diquark-antidiquark state strongly couples to the portion of the four-quark momentum space wave function where the relative momentum between the quasiparticles $\delta \equiv (Q_1 Q_3)$ and $\bar{\delta} \equiv \bar{Q}_2 \bar{Q}_4$ is considerably larger than the relative momenta within them. This dynamical diquark picture \cite{Brodsky:2014xia} was developed to provide a mechanism through which $\delta$-$\bar{\delta}$ states could live long enough to be identified as such experimentally. Later, this picture was developed into the dynamical diquark $\textit{model}$ \cite{Lebed:2017min} in which the separated $\delta$-$\bar{\delta}$ pair is connected by a color-flux-tube, whose quantized states are described well in terms of the various potentials computed using the Born-Oppenheimer (BO) approximation. The potentials used in that work are the potentials used in QCD lattice gauge-theory simulations that predict the spectrum of heavy-quarkonium hybrid mesons \cite{Juge:1997nc,Juge:1999ie,Juge:2002br,Morningstar:2019,Capitani:2018rox}. In Ref. \cite{Giron:2019bcs}, BO potentials are used in coupled Schrödinger equations and $\delta$-$\bar{\delta}$ spectrum was predicted. The mass spectrum and preferred decay modes with using of eigenstates of heavy-quark spin of the 6 isosinglets and 6 isotriplets were studied in Ref. \cite{Giron:2019cfc}. The dynamical diquark model was expanded to determine the mass spectra of states in the $P$-wave $c\bar{c}q\bar{q}^\prime$, $b\bar{b}q\bar{q}^\prime$, $c\bar{c}s\bar{s}$, $c\bar{c}c\bar{c}$, $c\bar{c}s\bar{q}$ and pentaquark sectors \cite{Giron:2020fvd,Giron:2020qpb,Giron:2020wpx,Giron:2021sla,Giron:2021fnl}, respectively, as well as radiative decay widths to $X(3872)$ of the lightest negative-parity exotic candidates \cite{Gens:2021wyf}.

If the fully-heavy  $cc\bar{c}\bar{c}$ and $bb\bar{b}\bar{b}$ tetraquark states do exist in nature, it is natural to believe that there exist more other heavy-flavor tetraquark systems. The mass spectra of $bc\bar{c}\bar{c}$ and $bc\bar{b}\bar{b}$ tetraquark states were studied among with other configurations \cite{Wu:2016vtq,Deng:2020iqw,Faustov:2020qfm,Liu:2019zuc}. Comparing to fully-heavy tetraquark systems, the $cc\bar{b}\bar{b}$ tetraquark states are quite different and interesting since they are doubly-charged and have no annihilation channels. They are expected to be stable if they lie below the $2B_c$ threshold. The ground state mass values of $cc\bar{b}\bar{b}$ tetraquarks were calculated  in a nonrelativistic effective field theory with OGE color Coulomb interaction and in a relativized diquark model characterized by OGE plus a confining potential \cite{Anwar:2017toa}. They gave an upper limit for mass of the $cc\bar{b}\bar{b}$ tetraquark as 12.58 GeV which is below the $2B_c$ threshold indicating that $cc\bar{b}\bar{b}$ tetraquark is stable against strong decays. The possibility of being stable of this state was not supported by the investigations of chromomagnetic interaction model \cite{Wu:2016vtq},  color-magnetic interaction model, a traditional constituent quark model and a multiquark color flux-tube model \cite{Deng:2020iqw}, relativistic quark model \cite{Faustov:2020qfm}, potential model \cite{Liu:2019zuc}, constituent quark model \cite{Czarnecki:2017vco}, nonrelativistic chiral quark model \cite{Chen:2019vrj}, nonrelativistic quark model \cite{Wang:2019rdo} and relativized diquark Hamiltonian \cite{Bedolla:2019zwg}. However, a recent study which was carried out by QCD moment sum rule method suggested that for the ground states of positive parity $cc\bar{b}\bar{b}$ tetraquarks are lower than $B_cB_c$ threshold and expected to be stable and narrow \cite{Wang:2021taf}. The outcomes of these works are quite different from each other. 

The mass spectroscopy has been a major effort to probe the dynamics of the tetraquarks.  To obtain the possible mass locations around the corresponding thresholds is not only  crucial for understanding the underlying dynamics of the exotic hadrons and the nature of strong interactions of QCD but also useful for experimental researches for their existence.

In this work, we systematically study the mass spectra of $cc\bar{b}\bar{b}$, $bc\bar{c}\bar{c}$, and $bc\bar{b}\bar{b}$ systems by using a dynamical diquark model where the exotic systems are constructed from heavy diquark $\delta$ and antidiquark $\bar{\delta}$. The paper is organized as follows. In Sec. \ref{sec:level2}, we describe the nomenclature of exotic tetraquarks in the dynamical diquark model. Sec. \ref{sec:level3} introduces the model of this work.  In Sec. \ref{sec:level3}, the $S-$wave mass spectra of the $cc\bar{b}\bar{b}$, $bc\bar{c}\bar{c}$, and $bc\bar{b}\bar{b}$ are calculated and confronted with the lowest thresholds for the fall-apart decays into two-meson. Sec. \ref{sec:level5} is a conclusion and summary of our results. 

\section{\label{sec:level2}Nomenclature of Exotic States}
The notion of diquarks dates back to the early days of the quark model. The concept of diquarks as effective degrees-of-freedom in quark models has been proven useful in the calculation of hadron spectra, see for an old but instructive review \cite{Anselmino:1992vg} and recent review \cite{Barabanov:2020jvn} and references therein. A diquark is defined as a colored bound state of two quarks. Up to now no diquark state have been detected in the experiments. However, experimental lack of observation does not eliminate the hypothesis of diquarks as constituents of tetraquarks and also baryons. 

The spectroscopy of $\delta-\bar{\delta}$ states was studied in which the diquarks with no internal orbital momentum but allowing for arbitrary orbital excitation and gluon-field excitation between the diquark pairs $\delta-\bar{\delta}$ in Ref. \cite{Lebed:2017min}. The core states which are expressed in the basis of good diquark-spin eigenvalues with labels such as $1_{\delta}$ are given by

\begin{eqnarray}
J^{PC} = 0^{++}: & \ & X_0 \equiv \vert 0_{\delta} , 0_{\bar{\delta}}\rangle_0 \, , \ \ X_0^\prime \equiv \vert 1_{\delta} , 1_{\bar{\delta}} \rangle_0, \nonumber \\
J^{PC} = 1^{++}: & \ & X_1 \equiv \frac{1}{\sqrt 2} \left(
\vert 1_{\delta} , 0_{\bar{\delta}}\rangle_1  + \vert 0_{\delta} , 1_{\bar{\delta}}\rangle_1 \right) , \nonumber \\
J^{PC} = 1^{+-}: & \ & Z \ \equiv
\frac{1}{\sqrt 2} \left(
\vert 1_{\delta} , 0_{\bar{\delta}}\rangle_1  - \vert 0_{\delta} , 1_{\bar{\delta}}\rangle_1 \right) , \nonumber \\ & \ & Z^\prime \,
\equiv \vert 1_{\delta} , 1_{\bar{\delta}} \rangle_1 \, , \nonumber \\
J^{PC} = 2^{++}: & \ & X_2 \equiv \vert 1_{\delta} , 1_{\bar{\delta}} \rangle_2 \ ,
\label{diquarkstates}
\end{eqnarray}
with the outer subscripts on the kets indicate total quark spin $S=J$ when orbital angular momentum is zero. These six states fill the lowest multiplet $\Sigma^+_g(1S)$ within the Born-Oppenheimer (BO) approximation for the gluon-field potential connecting the $\delta$-$\bar{\delta}$ pair. Note that the pairs $X_0$, $X_0^\prime$ and $Z$, $Z^\prime$ carry the same $J^{PC}$ quantum numbers and can certainly mix. One can define equivalent eigenstates \cite{Giron:2020qpb} which are $X_1$, $X_2$, and 

\begin{eqnarray}
{\tilde X}_0 & \equiv & \vert 0_{Q\bar{Q}}
 , 0_{Q^\prime \bar{Q}^\prime} \rangle_0 =
+ \frac{1}{2} X_0 + \frac{\sqrt{3}}{2} X_0^\prime \, , \nonumber \\
{\tilde X}_0^\prime & \equiv & \vert 1_{Q\bar{Q}} , 1_{Q^\prime \bar{Q}^\prime} \rangle_0 =
+ \frac{\sqrt{3}}{2} X_0 - \frac{1}{2} X_0^\prime \, , \nonumber \\
{\tilde Z} & \equiv & \vert 1_{Q\bar{Q}} , 0_{Q^\prime \bar{Q}^\prime} \rangle_1 =
\frac{1}{\sqrt{2}} \left( Z^\prime \! + Z \right) \, , \nonumber \\
{\tilde Z}^\prime & \equiv & \vert 0_{Q\bar{Q}} , 1_{Q^\prime \bar{Q}^\prime} \rangle_1 =
\frac{1}{\sqrt{2}} \left( Z^\prime \! - Z \right) \, .
\label{diquarkstatesmix}
\end{eqnarray}

For our purposes, we assume that the tetraquark states are colour antitriple ($\bar{3}_c$) diquark + colour triplet ($3_c$) antidiquark states. In this assumption, $X_0^\prime$, $Z^\prime$ and $X_2$ are the eigenstates that retain and we don't take care of the mixing effects which are obvious in Eq. (\ref{diquarkstatesmix}). $bb\bar{c}\bar{c}$, $bc\bar{c}\bar{c}$ and $bc\bar{b}\bar{b}$ states do not have determined $C$- parity. Therefore it is not a ``good" quantum number for these states.

\section{\label{sec:level3}Numerical Aspects of Born-Oppenheimer Potential}
In the dynamical diquark model, the mass spectrum of all states is computed by a particular BO potential, $\Gamma$ ($\Sigma_g^+$, $\Sigma_u^+$, \emph{etc}.) which gives rise to a multiplet of states. These potentials are calculated on the lattice by several groups~\cite{Juge:1997nc,Juge:1999ie,Juge:2002br,Morningstar:2019,Capitani:2018rox}. Then, using only the mass of a diquark $m_{\delta}$ and antidiquark $m_{\bar{\delta}}$, one can solve the resulting Schr\"{o}dinger equations for this Hamiltonian numerically~\cite{Giron:2019cfc}. In some cases, the BO potentials can mix leading to coupled Schr\"{o}dinger equations whose numerical solutions involve techniques outlined in Ref.~\cite{Giron:2019cfc}. Once one has a solution to this Hamiltonian, this gives rise to a multiplet-average mass eigenvalue $M_0$ which is unique to a particular radial ($n$) and orbital ($L$) quantum number attached to the particular BO potential. More in depth details about the specific BO potentials can be found in Ref.~\cite{Lebed:2017min}. 

The governing equations for the dynamical diquark model can be found in detail in Ref.~\cite{Giron:2019bcs} and will be highlighted briefly in the following section. The radial Schr\"{o}dinger equations for the uncoupled BO potentials $V_\Gamma(r)$ can be written as, in natural units for reduced mass $\mu$, 
\begin{equation}
\left[-\frac{1}{2\mu r^2}\partial_r r^2 \partial_r + \frac{\ell\left(\ell +1\right)}{2\mu r^2}+V_\Gamma(r)\right]\psi_\Gamma^{(n)}(r)=E_n\psi_\Gamma^{(n)},
\end{equation}
and the coupled BO potentials read:
\begin{eqnarray}
&&\left[-\frac{1}{2\mu r^2}\partial_r r^2 \partial_r +\frac{1}{2\mu r^2}
\begin{pmatrix}
\ell\left(\ell+1\right) +2 & 2\sqrt{\ell\left(\ell+1\right)}\\
2\sqrt{\ell\left(\ell+1\right)} & \ell\left(\ell+1\right)
\end{pmatrix}\right.\nonumber\\
&& \left. 
\begin{pmatrix}
V_{\Sigma_u^-} & 0\\
0 & V_{\Pi_u^+}
\end{pmatrix}\right]
\begin{pmatrix}
\psi_{\Sigma_u^-}^{(n)}(r)\\
\psi_{\Pi_u^+}^{(n)}(r)
\end{pmatrix}= E_n
\begin{pmatrix}
\psi_{\Sigma_u^-}^{(n)}(r)\\
\psi_{\Pi_u^+}^{(n)}(r)
\end{pmatrix}.
\end{eqnarray}

We use the functional form of $V(r)$ given by lattice simulations \cite{Juge:2002br,Morningstar:2019}. The $S-$wave Hamiltonian $QQ\bar{Q}\bar{Q}$ can be written as
\begin{eqnarray} 
H&=& H_0 + 2\left[\kappa_{QQ}\left(\bold{s}_Q \cdot \bm{s}_Q \right)+
\kappa_{\bar{Q}\bar{Q}} \left(\bold{s}_{\bar{Q}} \cdot \bm{s}_{\bar{Q}}\right)
\right].
\label{Hamiltonian}
\end{eqnarray}
The eigenvalues of $H$ are computed as
\begin{equation}
M= M_0 + \Delta M_{\kappa_{QQ}} + \Delta M_{\kappa_{\bar{Q}\bar{Q}}},
\end{equation}
where 
\begin{equation}
\Delta M_{\kappa_{QQ}} =\frac 1 2 \kappa_{QQ}\left[2s_\delta \left(s_\delta +1\right)-3
\right],
\end{equation}
and
\begin{equation}
\Delta M_{\kappa_{\bar{Q}\bar{Q}}}= \frac 1 2 \kappa_{\bar{Q}\bar{Q}}
\left[2s_{\bar{\delta}} \left(s_{\bar{\delta}} +1\right) -3 \right].
\end{equation}
In the above equations, $s_\delta$ represents the spin of the diquark and $s_{\bar{\delta}}$ represents the spin of the antidiquark. 

\section{\label{sec:level4}Numerical Results and Discussion}
We prefer to make a prediction that is independent of the nonrelativistic quark models. This was also done in Ref. \cite{Brodsky:2014xia}. So we use the mass values of the for $J^P=1^+$ axial vector states $cc$, $\bar{b}\bar{b}$ and $cb$ diquarks as
\begin{eqnarray}
m_{cc}&=& 3.51 ~ \text{GeV}, \nonumber \\
m_{bb}&=& 8.67 ~ \text{GeV}, \nonumber  \\
m_{cb}&=& 5.80 ~ \text{GeV}, 
\label{masssr}
\end{eqnarray}
where $m_{cc}$ and $m_{bb}$ are taken from Ref. \cite{Esau:2019hqw} whereas $m_{cb}$ is taken from Ref. \cite{Tang:2011fv}. The coupling constants $\kappa$ were assumed to be $\kappa = 0< \kappa_{cc},\kappa_{cb}, \kappa_{bb} < 67 ~ \text{MeV}$ in the same work. Since the spin couplings $\kappa_{cc}$, $\kappa_{bc}$, and $\kappa_{bb}$ are quite distinct and expected to decrease with increasing quark masses \cite{Maiani:2004vq}, we choose two sets of parameters for calculations. In the first set we choose $\kappa_{cc}=60 ~ \text{MeV}$, $\kappa_{cb}=50 ~ \text{MeV}$, $\kappa_{bb}=40~ \text{MeV}$ whereas in the second set we choose $\kappa_{cc}=50 ~ \text{MeV}$, $\kappa_{cb}=40 ~ \text{MeV}$,  $\kappa_{bb}=30~ \text{MeV}$. The value of $\kappa_{cc}$ is different from the Ref. \cite{Giron:2020wpx} because 2S/1P states of $c\bar{c}c\bar{c}$ were taken into account. 
 
\subsection{$cc\bar{b}\bar{b}$ State}
The mass spectra and possible $S$-wave thresholds of $cc\bar{b}\bar{b}$ states are listed in Table \ref{ccbbmass}. The meson masses are taken from  Ref. \cite{Zyla:2020zbs} and the predicted mass for $M_{B_c^\ast}=6388$ MeV is barrowed from \cite{Godfrey:1985xj}. We also show values of the difference of the fully-heavy tetraquark and threshold masses, $\Delta=M-T$. If this quantity is negative, then the fully-heavy tetraquark state lies below the threshold of the fall-apart decay into two mesons and thus should be a narrow state. The states with small positive values of $\Delta$ could be also observed as resonances, since their decay rates will be suppressed by the phase space. All other states are expected to be broad and thus difficult to observe in the experiments. 

\begin{table}[H]
\begin{center}
\caption{\label{ccbbmass} Predicted mass spectra for the $cc\bar{b}\bar{b}$ states, corresponding thresholds and $\Delta$ values. The superscripts denote the spin of the diquark/antidiquark whereas the subscripts denote the color structure. $\{~\}$ denotes the symmetric flavor wave functions of the two quarks and antiquarks subsystems. All results are in units of MeV.}
\begin{tabular*}{18cm}{@{\extracolsep{\fill}}p{1cm}<{\centering}p{1cm}<{\centering}p{1cm}<{\centering}p{4cm}<{\centering}p{3cm}<{\centering}}
\hline\hline
 $J^{P}$  & Configuration  & Mass &$S$-wave threshold &  $\Delta$ \\ \hline
Set I & && & \\  \hline
 $0^{+}$  & $\left[\{cc\}^1_{\bar{\mathbf{3}}_c}\{\bar{b}\bar{b}\}^1_{\mathbf{3}_c}\right]^0_{\mathbf{1}_c}$ & \multirow{1}{*} {$\begin{bmatrix} 12401 \end{bmatrix}$} &\multirow{1}{*}{$\begin{bmatrix}(B_c^{\pm}B_c^{\pm}, B_c^{\ast\pm}B_c^{\ast\pm})  \end{bmatrix}$} & \multirow{1}{*}{$\begin{bmatrix}(-148, -375) \end{bmatrix}$} \\
            
 $1^{+}$  & $\left[\{cc\}^1_{\bar{\mathbf{3}}_c}\{\bar{b}\bar{b}\}^1_{\mathbf{3}_c}\right]^1_{\mathbf{1}_c}$     & \multirow{1}{*} {$\begin{bmatrix} 12409 \end{bmatrix}$}
               & \multirow{1}{*}{$\begin{bmatrix}(B_c^{\pm}B_c^{\ast\pm}, B_c^{\ast\pm}B_c^{\ast\pm})  \end{bmatrix}$} & \multirow{1}{*}{$\begin{bmatrix}(-253, -367) \end{bmatrix}$}\\
                
 $2^{+}$  &  $\left[\{cc\}^1_{\bar{\mathbf{3}}_c}\{\bar{b}\bar{b}\}^1_{\mathbf{3}_c}\right]^2_{\mathbf{1}_c}$    &\multirow{1}{*} {$\begin{bmatrix} 12427 \end{bmatrix}$}    & \multirow{1}{*}{$\begin{bmatrix} B_c^{\ast\pm}B_c^{\ast\pm}  \end{bmatrix}$}  & \multirow{1}{*} {$\begin{bmatrix}  -349 \end{bmatrix}$} \\ \hline
 
 Set II & && & \\  \hline
 $0^{+}$  & $\left[\{cc\}^1_{\bar{\mathbf{3}}_c}\{\bar{b}\bar{b}\}^1_{\mathbf{3}_c}\right]^0_{\mathbf{1}_c}$ & \multirow{1}{*} {$\begin{bmatrix} 12381 \end{bmatrix}$} &\multirow{1}{*}{$\begin{bmatrix}(B_c^{\pm}B_c^{\pm}, B_c^{\ast\pm}B_c^{\ast\pm})  \end{bmatrix}$} & \multirow{1}{*}{$\begin{bmatrix}(-168, -395) \end{bmatrix}$} \\
            
 $1^{+}$  & $\left[\{cc\}^1_{\bar{\mathbf{3}}_c}\{\bar{b}\bar{b}\}^1_{\mathbf{3}_c}\right]^1_{\mathbf{1}_c}$     & \multirow{1}{*} {$\begin{bmatrix} 12390 \end{bmatrix}$}
               & \multirow{1}{*}{$\begin{bmatrix}(B_c^{\pm}B_c^{\ast\pm}, B_c^{\ast\pm}B_c^{\ast\pm})  \end{bmatrix}$} & \multirow{1}{*}{$\begin{bmatrix}(-272, -386) \end{bmatrix}$}\\
                
 $2^{+}$  &  $\left[\{cc\}^1_{\bar{\mathbf{3}}_c}\{\bar{b}\bar{b}\}^1_{\mathbf{3}_c}\right]^2_{\mathbf{1}_c}$    &\multirow{1}{*} {$\begin{bmatrix} 12408 \end{bmatrix}$}    & \multirow{1}{*}{$\begin{bmatrix} B_c^{\ast\pm}B_c^{\ast\pm}  \end{bmatrix}$}  & \multirow{1}{*} {$\begin{bmatrix}  -368\end{bmatrix}$} \\
 
\hline\hline
\end{tabular*}
\end{center}
\end{table}

All the predicted $cc\bar{b}\bar{b}$ mass values of the present model are lower than their lowest open flavor decay channels. This implies that these tetraquark states can only undergo radiative transitions or weak decays. According to model of this work, they are expected to be very narrow and stable if they do exist. Among these decay states, $B_cB_c^\ast$ and $B_c^\ast B_c^\ast$ are not possible in near future since only the ground state of $B_c$ meson was confirmed in the experiments. When enough data for $B_c^\ast$ are obtained, the corresponding decay channels can be searched. 

We compare our results with the available studies in the literature. The results are presented in Table \ref{comparison}. The mass splitting is 26 MeV for set I and 27 MeV for set II. This small mass splitting may be due to the both small $\kappa$ diquark internal spin-spin couplings. Accordingly, mass splittings are 98 MeV in color-magnetic interaction model (CMIM), 54 MeV in multiquark color flux-tube model (MCFTM) and 78 MeV in traditional constituent quark model (CQM) of Ref. \cite{Deng:2020iqw}, 79 Mev for Ref. \cite{Liu:2019zuc}, 21 MeV for Model Ia, 4 MeV for Model Ib, 54 MeV for Model IIa and 16 MeV for Model IIb of Ref. \cite{Wang:2019rdo} and 169 MeV for Ref. \cite{Bedolla:2019zwg}. 

The predicted mass values of $0^{+}$ agree well with the result of Ref. \cite{Wang:2021taf} in which $\bar{3}_c \otimes 3_c$  type interpolating current is used and result of Ref. \cite{Bedolla:2019zwg} where relativized diquark model was used. It is interesting to note that two of pure $0^+$ $6_c \otimes \bar{6}_c$ type interpolating currents yielded mass results that are in good agreement with our predictions. Our results differ at the order of 500 MeV comparing to other references. It should be mentioned that some references take into account mixing of configurations in $0^{+}$. For $1^{+}$, predicted mass values of our work agree well with the result of $\bar{3}_c \otimes 3_c$ type interpolating current of Ref. \cite{Wang:2021taf}. In the $2^{+}$, our results are compatible with the result of Ref. \cite{Wang:2021taf} where $\bar{3}_c \otimes 3_c$ type interpolating current was used. Our results differ at the order of 600 MeV in both $1^{+}$ and $2^{+}$ quantum numbers except $6_c \otimes \bar{6}_c$ states. The differences among reference studies arise from the different choices of interactions, parameters, or even the models that are included and used. 

\begin{table}[H]
\caption{Comparison of theoretical predictions for the masses of the $cc\bar{b}\bar{b}$  tetraquark states. All results are in units of MeV except Ref. \cite{Wang:2021taf} where results are in units of GeV.}
 \label{comparison}
\begin{ruledtabular}
\begin{tabular}{ccccccccccc}
\cline{2-3} 
Reference & \multicolumn{2}{c}{$0^{+}$} & $1^{+}$ & $2^{+}$\\
\hline
& $\bar{3}_c \otimes 3_c$ & $6_c \otimes \bar{6}_c$ & &  \\ \hline
Set I & 12401 & -  & 12409 & 12427 \\
Set II & 12381 & -  & 12390 & 12408 \\
\cite{Deng:2020iqw} (CMIM) & 12597 $(66\%)$& 12597 $(34\%)$ &12660 &12695 \\ 
\cite{Deng:2020iqw} (MCFTM) & 12906 $(49\%)$& 12906 $(51\%)$ &12945 &12960 \\ 
\cite{Deng:2020iqw} (CQM) & 12963 $(29\%)$& 12963 $(71\%)$ &13024 &13041 \\ 
\cite{Liu:2019zuc} (PM)& 12953 & 13032 & 12960 & 12972 \\
\cite{Wang:2019rdo} Mod. Ia (NRQM) & 12847 $(14 \%)$& 12847 $(86 \%)$ &12864 &12868  \\
\cite{Wang:2019rdo} Mod. Ib (NRQM)& 12866 $(86\%)$& 12866 $(14\%)$ &12864 &12868  \\
\cite{Wang:2019rdo} Mod. IIa (NRQM)& 12886 $(53\%)$& 12886 $(47\%)$ &12924 &12940  \\
\cite{Wang:2019rdo} Mod. IIb (NRQM)& 12946 $(47\%)$& 12946 $(53\%)$ &12924 &12940  \\
\cite{Bedolla:2019zwg} (RDH) & 12445 &- &12536 &12614 \\
\cite{Wang:2021taf} (QCDSR) & & $13.32^{+0.30}_{-0.24}$ & &\\ 
\cite{Wang:2021taf} (QCDSR) & & $12.41^{+0.21}_{-0.17}$ & &\\ 
\cite{Wang:2021taf} (QCDSR) & & $12.36^{+0.18}_{-0.15}$ & &\\ 
\cite{Wang:2021taf} (QCDSR) & $12.33^{+0.18}_{-0.15}$&  & &\\ 
\cite{Wang:2021taf} (QCDSR) & $12.36^{+0.19}_{-0.16}$&  & &\\ 
\cite{Wang:2021taf} (QCDSR) & &  & $13.35^{+0.33}_{-0.26}$ $(6_c \otimes \bar{6}_c)$&\\ 
\cite{Wang:2021taf} (QCDSR) & &  &$13.33^{+0.28}_{-0.22}$ $(6_c \otimes \bar{6}_c)$ &\\ 
\cite{Wang:2021taf} (QCDSR) & &  &$12.36^{+0.19}_{-0.16}$ $(\bar{3}_c \otimes 3_c)$ &\\ 
\cite{Wang:2021taf} (QCDSR) & &  &$12.34^{+0.18}_{-0.15}$ $(\bar{3}_c \otimes 3_c)$ &\\ 
\cite{Wang:2021taf} (QCDSR) & &  & &$13.41^{+0.34}_{-0.26}$ $(6_c \otimes \bar{6}_c)$\\ 
\cite{Wang:2021taf} (QCDSR) & &  & &$12.27^{+0.19}_{-0.16}$ $(\bar{3}_c \otimes 3_c)$\\ 
\end{tabular}
\end{ruledtabular}
\end{table}

We also calculate root mean square $\sqrt{\langle r^2 \rangle}$ between the $cc$ and $\bar{b}\bar{b}$ in the $cc\bar{b}\bar{b}$ state and compare with the available studies in Table \ref{rms}. The obtained results of our model agree well with the results of references. We observe from the table that $cc\bar{b}\bar{b}$ tetraquark is compact. 

\begin{table}[H]
\caption{Comparison of the ground state root mean square radii $\sqrt{\langle r^2 \rangle} $ between the $cc$ and $\bar{b}\bar{b}$ in the $cc\bar{b}\bar{b}$ state. All results are in units of fm.}
 \label{rms}
\begin{ruledtabular}
\begin{tabular}{ccccccccccc}
\cline{2-4} 
Reference & \multicolumn{3}{c}{$0^{+}$} & $1^{+}$ & $2^{+}$\\
\hline
& $\bar{3}_c \otimes 3_c$ & &$6_c \otimes \bar{6}_c$ & &  \\ \hline
This work &  &0.24  & -  & 0.25 & 0.27 \\
\cite{Deng:2020iqw} (MCFTM)&49\% &0.25 &51\% &0.29 & 0.30 \\
\cite{Wang:2019rdo} Model I & 14\% &0.26& 86 \% &- &- \\
\cite{Wang:2019rdo} Model II & 53\% &0.26& 47 \% &- &- 
\end{tabular}
\end{ruledtabular}
\end{table}

\subsection{$bc\bar{c}\bar{c}$ State}
We present the results for $bc\bar{c}\bar{c}$ tetraquark state in Table \ref{bcccmass}. As can be seen, all the predicted masses are above the corresponding $S$-wave thresholds. Therefore these states should be broad and are difficult to observe in experiments. The mass splitting is 34 MeV for both set I and II. We compare our results with the available studies in the literature. The results are presented in Table \ref{comparison2}. As can be seen in Table \ref{comparison2}, our results are compatible with the results of Ref. \cite{Faustov:2020qfm} in which the authors used a relativistic diquark-antidiquark model. Our results differ approximately 180 MeV comparing to results of Ref. \cite{Liu:2019zuc} and approximately 250 MeV with respect to results of Ref. \cite{Deng:2020iqw}. The mass splittings are 128 MeV for (CMIM), 62 MeV for (MCFTM), 86 MeV for (CQM) of Ref. \cite{Deng:2020iqw}, 75 MeV for Ref. \cite{Faustov:2020qfm} and 28 MeV for Ref. \cite{Liu:2019zuc}. We present root mean square radii of $bc\bar{c}\bar{c}$ state in Table \ref{rms2}. The results show that $bc\bar{c}\bar{c}$ state is compact.

\begin{table}[H]
\begin{center}
\caption{\label{bcccmass} Predicted mass spectra for the $bc\bar{c}\bar{c}$ states, corresponding thresholds and $\Delta$ values. $\{~\}$ denotes the symmetric flavor wave functions of the two quarks (antiquarks) and $()$ denotes no permutation symmetries in these quark pairs. All results are in units of MeV.}
\begin{tabular*}{18cm}{@{\extracolsep{\fill}}p{1cm}<{\centering}p{1cm}<{\centering}p{1cm}<{\centering}p{4cm}<{\centering}p{3cm}<{\centering}}
\hline\hline
 $J^{P}$  & Configuration  & Mass &  $S$-wave threshold &  $\Delta$ \\ \hline
Set I & && & \\  \hline
 $0^{+}$  & $\left[(bc)^1_{\bar{\mathbf{3}}_c}\{\bar{c}\bar{c}\}^1_{\mathbf{3}_c}\right]^0_{\mathbf{1}_c}$ & \multirow{1}{*} {$\begin{bmatrix} 9579 \end{bmatrix}$} &\multirow{1}{*}{$\begin{bmatrix}(\eta_c B_c^{\pm}, J/\psi B_c^{\ast\pm})  \end{bmatrix}$} & \multirow{1}{*}{$\begin{bmatrix}(321, 94) \end{bmatrix}$} \\
            
 $1^{+}$  & $\left[(bc)^1_{\bar{\mathbf{3}}_c}\{\bar{c}\bar{c}\}^1_{\mathbf{3}_c}\right]^1_{\mathbf{1}_c}$     & \multirow{1}{*} {$\begin{bmatrix} 9590 \end{bmatrix}$}
               & \multirow{1}{*}{$\begin{bmatrix}(\eta_c B_c^{\ast\pm},J/\psi B_c^{\pm},J/\psi B_c^{\ast\pm})  \end{bmatrix}$} & \multirow{1}{*}{$\begin{bmatrix}(218,219,105) \end{bmatrix}$}\\
                
 $2^{+}$  &  $\left[(bc)^1_{\bar{\mathbf{3}}_c}\{\bar{c}\bar{c}\}^1_{\mathbf{3}_c}\right]^2_{\mathbf{1}_c}$    &\multirow{1}{*} {$\begin{bmatrix} 9613 \end{bmatrix}$}    & \multirow{1}{*}{$\begin{bmatrix} J/\psi B_c^{\ast\pm} \end{bmatrix}$}  & \multirow{1}{*} {$\begin{bmatrix} 128 \end{bmatrix}$} \\ \hline
 
 Set II & && & \\  \hline
 $0^{+}$  & $\left[(bc)^1_{\bar{\mathbf{3}}_c}\{\bar{c}\bar{c}\}^1_{\mathbf{3}_c}\right]^0_{\mathbf{1}_c}$ & \multirow{1}{*} {$\begin{bmatrix} 9560  \end{bmatrix}$} &\multirow{1}{*}{$\begin{bmatrix}(\eta_c B_c^{\pm}, J/\psi B_c^{\ast\pm}) \end{bmatrix}$} & \multirow{1}{*}{$\begin{bmatrix}(302, 75) \end{bmatrix}$} \\
            
 $1^{+}$  & $\left[(bc)^1_{\bar{\mathbf{3}}_c}\{\bar{c}\bar{c}\}^1_{\mathbf{3}_c}\right]^1_{\mathbf{1}_c}$     & \multirow{1}{*} {$\begin{bmatrix} 9571  \end{bmatrix}$}
               & \multirow{1}{*}{$\begin{bmatrix}(\eta_c B_c^{\ast\pm},J/\psi B_c^{\pm},J/\psi B_c^{\ast\pm})  \end{bmatrix}$} & \multirow{1}{*}{$\begin{bmatrix}(199,200, 86) \end{bmatrix}$}\\
                
 $2^{+}$  &  $\left[(bc)^1_{\bar{\mathbf{3}}_c}\{\bar{c}\bar{c}\}^1_{\mathbf{3}_c}\right]^2_{\mathbf{1}_c}$    &\multirow{1}{*} {$\begin{bmatrix} 9594  \end{bmatrix}$}    & \multirow{1}{*}{$\begin{bmatrix} J/\psi B_c^{\ast\pm}  \end{bmatrix}$}  & \multirow{1}{*} {$\begin{bmatrix} 109 \end{bmatrix}$} \\
 
\hline\hline
\end{tabular*}
\end{center}
\end{table}

\begin{table}[H]
\caption{Comparison of theoretical predictions for the masses of the $bc\bar{c}\bar{c}$  tetraquark states.  All results are in units of MeV.}
 \label{comparison2}
\begin{ruledtabular}
\begin{tabular}{ccccccccccc}
\cline{2-3} 
\cline{4-5}
\cline{6-7}
Reference & \multicolumn{2}{c}{$0^{+}$} & \multicolumn{2}{c} {$1^{+}$} & \multicolumn{2}{c} {$2^{+}$}\\
\hline
& $\bar{3}_c \otimes 3_c$ & $6_c \otimes \bar{6}_c$ & $\bar{3}_c \otimes 3_c$ & $6_c \otimes \bar{6}_c$ &$\bar{3}_c \otimes 3_c$ & $6_c \otimes \bar{6}_c$   \\ \hline
Set I & 9579 & -  & 9590 &- &9613 &- \\
Set II & 9560 & -  & 9571 &- &9594 &- \\
\cite{Deng:2020iqw} (CMIM) & 9314 $(66\%)$& 9314 $(34\%)$ &9343 $(65\%)$  &9343 $(35\%) $ & 9442 $(100\%)$ &-\\ 
\cite{Deng:2020iqw} (MCFTM) & 9670 $(56\%)$& 9670 $(44\%)$ &9683 $(58\%)$  &9683 $(42\%) $ & 9732 $(100\%)$ &-\\ 
\cite{Deng:2020iqw} (CQM) & 9753 $(41\%)$& 9753 $(59\%)$ &9766 $(31\%)$  &9766 $(59\%) $ & 9839 $(100\%)$ &-\\ 
\cite{Faustov:2020qfm} & 9572 & - &9602  &-  & 9647 &-\\ 
\cite{Liu:2019zuc} & 9740 & - &9749  &-  & 9768 &-\\ 
\end{tabular}
\end{ruledtabular}
\end{table}

\begin{table}[H]
\caption{Ground state root mean square radii $\sqrt{\langle r^2 \rangle} $ between the $bc$ and $\bar{c}\bar{c}$ in the $bc\bar{c}\bar{c}$ state. All results are in units of fm.}
 \label{rms2}
\begin{ruledtabular}
\begin{tabular}{cccccc}
 & $0^{+}$ & $1^{+}$ & $2^{+}$\\
\hline
This work & 0.23 &  0.24 & 0.24 \\
\end{tabular}
\end{ruledtabular}
\end{table}

\subsection{$bc\bar{b}\bar{b}$ State}

We present the results for $bc\bar{c}\bar{c}$ tetraquark state in Table \ref{bcbbmass}. As can be seen, all the predicted masses are at most 387 MeV above the corresponding $S$-wave thresholds. Therefore, these states should be broad and are difficult to observe in experiments. The mass splitting is 8 MeV for set I and 9 MeV for set II. We compare our results with the available studies in the literature in Table \ref{comparison3}. As can be seen in Table \ref{comparison3}, our results are in compatible with the reference studies except (CMIM) of Ref. \cite{Deng:2020iqw}. The mass splittings are 93 MeV for (CMIM), 56 MeV for (MCFTM), 99 MeV for (CQM) of Ref. \cite{Deng:2020iqw}, 23 MeV for Ref. \cite{Faustov:2020qfm}, 18 MeV for Ref. \cite{Liu:2019zuc}. We present root mean square radii of $bc\bar{b}\bar{b}$ state in Table \ref{rms3}. The results show that $bc\bar{b}\bar{b}$ state is compact.

\begin{table}[H]
\begin{center}
\caption{\label{bcbbmass} Predicted mass spectra for the $bc\bar{b}\bar{b}$ states, corresponding thresholds and $\Delta$ values.  $\{~\}$ denotes the symmetric flavor wave functions of the two quarks (antiquarks) and $()$ denotes no permutation symmetries in these quark pairs. All results are in units of MeV.}
\begin{tabular*}{18cm}{@{\extracolsep{\fill}}p{1cm}<{\centering}p{1cm}<{\centering}p{1cm}<{\centering}p{4cm}<{\centering}p{3cm}<{\centering}}
\hline\hline
 $J^{P}$  & Configuration  & Mass &  $S$-wave threshold &  $\Delta$ \\ \hline
Set I & && & \\  \hline
 $0^{+}$  & $\left[(bc)^1_{\bar{\mathbf{3}}_c}\{\bar{b}\bar{b}\}^1_{\mathbf{3}_c}\right]^0_{\mathbf{1}_c}$ & \multirow{1}{*} {$\begin{bmatrix} 16060 \end{bmatrix}$} &\multirow{1}{*}{$\begin{bmatrix}(B_c^{\pm}\eta_b, B_c^{\ast\pm}\Upsilon)  \end{bmatrix}$} & \multirow{1}{*}{$\begin{bmatrix}(387, 212) \end{bmatrix}$} \\
            
 $1^{+}$  & $\left[(bc)^1_{\bar{\mathbf{3}}_c}\{\bar{b}\bar{b}\}^1_{\mathbf{3}_c}\right]^1_{\mathbf{1}_c}$     & \multirow{1}{*} {$\begin{bmatrix} 16062 \end{bmatrix}$}
               & \multirow{1}{*}{$\begin{bmatrix}(B_c^{\pm}\Upsilon,B_c^{\ast\pm}\eta_b ,B_c^{\ast\pm}\Upsilon)  \end{bmatrix}$} & \multirow{1}{*}{$\begin{bmatrix}(327,275,214) \end{bmatrix}$}\\
                
 $2^{+}$  &  $\left[(bc)^1_{\bar{\mathbf{3}}_c}\{\bar{b}\bar{b}\}^1_{\mathbf{3}_c}\right]^2_{\mathbf{1}_c}$    &\multirow{1}{*} {$\begin{bmatrix} 16068 \end{bmatrix}$}    & \multirow{1}{*}{$\begin{bmatrix} B_c^{\ast\pm}\Upsilon \end{bmatrix}$}  & \multirow{1}{*} {$\begin{bmatrix} 220 \end{bmatrix}$} \\ \hline
 
 Set II & && & \\  \hline
 $0^{+}$  & $\left[(bc)^1_{\bar{\mathbf{3}}_c}\{\bar{b}\bar{b}\}^1_{\mathbf{3}_c}\right]^0_{\mathbf{1}_c}$ & \multirow{1}{*} {$\begin{bmatrix} 16049 \end{bmatrix}$} &\multirow{1}{*}{$\begin{bmatrix}(B_c^{\pm}\eta_b, B_c^{\ast\pm}\Upsilon)   \end{bmatrix}$} & \multirow{1}{*}{$\begin{bmatrix}(376,201) \end{bmatrix}$} \\
            
 $1^{+}$  & $\left[(bc)^1_{\bar{\mathbf{3}}_c}\{\bar{b}\bar{b}\}^1_{\mathbf{3}_c}\right]^1_{\mathbf{1}_c}$     & \multirow{1}{*} {$\begin{bmatrix} 16052 \end{bmatrix}$}
               & \multirow{1}{*}{$\begin{bmatrix}(B_c^{\pm}\Upsilon,B_c^{\ast\pm}\eta_b ,B_c^{\ast\pm}\Upsilon)  \end{bmatrix}$} & \multirow{1}{*}{$\begin{bmatrix}(317,265,204) \end{bmatrix}$}\\
                
 $2^{+}$  &  $\left[(bc)^1_{\bar{\mathbf{3}}_c}\{\bar{b}\bar{b}\}^1_{\mathbf{3}_c}\right]^2_{\mathbf{1}_c}$    &\multirow{1}{*} {$\begin{bmatrix} 16058 \end{bmatrix}$}    & \multirow{1}{*}{$\begin{bmatrix}B_c^{\ast\pm}\Upsilon  \end{bmatrix}$}  & \multirow{1}{*} {$\begin{bmatrix} 210 \end{bmatrix}$} \\
 
\hline\hline
\end{tabular*}
\end{center}
\end{table}

\begin{table}[H]
\caption{\label{comparison3}Comparison of theoretical predictions for the masses of the $bc\bar{b}\bar{b}$ tetraquark states. All  results are in units of MeV.}
\begin{ruledtabular}
\begin{tabular}{ccccccccccc}
\cline{2-3} 
\cline{4-5}
\cline{6-7}
Reference & \multicolumn{2}{c}{$0^{+}$} & \multicolumn{2}{c} {$1^{+}$} & \multicolumn{2}{c} {$2^{+}$}\\
\hline
& $\bar{3}_c \otimes 3_c$ & $6_c \otimes \bar{6}_c$ & $\bar{3}_c \otimes 3_c$ & $6_c \otimes \bar{6}_c$ &$\bar{3}_c \otimes 3_c$ & $6_c \otimes \bar{6}_c$   \\ \hline
Set I & 16060 & -  & 16062  &- & 16068  &- \\
Set II & 16049 & -  & 16052 &- & 16058 &- \\
\cite{Deng:2020iqw} (CMIM) & 15713 $(66\%)$& 15713 $(34\%)$ &15729 $(67\%)$  &15729 $(43\%) $ & 15806 $(100\%)$ &-\\ 
\cite{Deng:2020iqw} (MCFTM) & 16126 $(42\%)$& 16126 $(58\%)$ &16130 $(39\%)$  &16130 $(61\%) $ & 16182 $(100\%)$ &-\\ 
\cite{Deng:2020iqw} (CQM) & 16175 $(31\%)$& 16175 $(69\%)$ &16179 $(23\%)$  &16179 $(77\%) $ & 16274 $(100\%)$ &-\\ 
\cite{Faustov:2020qfm} & 16109 & - & 16117  &-  & 16132 &-\\ 
\cite{Liu:2019zuc} & 16158 & - &16164  &-  & 16176 &-\\ 
\end{tabular}
\end{ruledtabular}
\end{table}

\begin{table}[H]
\caption{Ground state root mean square radii $\sqrt{\langle r^2 \rangle} $ between the $bc$ and $\bar{b}\bar{b}$ in the $bc\bar{b}\bar{b}$ state. All results are in units of fm.}
 \label{rms3}
\begin{ruledtabular}
\begin{tabular}{ccccccccccc}
 &$0^{+}$ & $1^{+}$ & $2^{+}$\\
\hline
This work & 0.26 & 0.27 & 0.27\\
\end{tabular}
\end{ruledtabular}
\end{table}

\subsection{Brief Discussion on Diquark Models}

It should be noted that there are different types of diquark models in the literature. In Ref. \cite{Maiani:2004vq}, the diquark model is defined in a Hamiltonian formalism, using as interaction operators local spin-spin couplings together with spin-orbit and purely orbital terms. This model treats tetraquarks as bound $(\delta \overline{\delta})$ molecules and take into account spin-spin couplings between the various component quarks. Ref. \cite{Maiani:2014aja} presents a diquark model by neglecting all spin-spin couplings between a quark of the diquark and an antiquark of the antidiquark. In this model, interactions are restricted to spin-spin couplings which are assumed to be occur between quarks within either the $\delta$ or the $\overline{\delta}$. 

Refs. \cite{Maiani:2019cwl,Maiani:2019lpu} introduced a new perspective of exotic hadrons and modified the diquark model including a short-range repulsion between diquarks. This new framework uses Born–Oppenheimer approximation to study QCD interactions of multiquark hadrons containing heavy and light quarks. This model consists in solving eigenvalue problem for the light particles with fixed coordinates of the heavy particles, and then solve Schr\"{o}dinger equation of the heavy particles in the BO potential. The fundamental idea behind this framework is that tetraquarks are considered in terms of color molecules: two lumps of two-quark (colored atoms) held together by color forces. They observed that exotic X and Z resonances may emerge as QCD molecular objects made of colored two-quark lumps in which heavy-light diquarks are spatially separated from antidiquarks. In Ref. \cite{Maiani:2019cwl}, it was shown for BO potential for $(QQ)_{\overline{3}}$ that forces among constituents are all attractive and the potential vanishes at large $(QQ)$ separation. Ref. \cite{Maiani:2019lpu} also pointed out that the $6_c − \overline{6}_c$ configuration is also very important to form the tetraquark states. It was observed that $q \bar q$ repulsion in $Q \bar Q q \bar q$ is critical parameter to determine the internal configuration of tetraquark. 

The separations between the diquarks are important tools in the diquark models. In the original diquark picture \cite{Brodsky:2014xia,Lebed:2017min}, the diquark separation emerge as a result of the production mechanism. As mentioned in Ref. \cite{Giron:2020wpx}, the diquark-antidiquark state couples mostly to the portion of the four-quark momentum space wave function. Here the relative momentum between the $\delta-\overline{\delta}$ is significantly larger than the relative momenta within them. Alternatively, this separation can also be obtained by imposing a potential barrier \cite{Maiani:2019cwl}.

The Born-Oppenheimer approximation requires at least two-heavy sources plus light degree of freedoms for four-quark systems. In heavy-light $QQ \bar q \bar q$ tetraquarks, mass of the heavy quark is much bigger than the mass of the light quark, $M_Q >> m_q$. Relativistic motion of light quarks in the field of the heavy color sources produces an effective potential which is a function of the relative distance $R$ separating the heavy quark $QQ$ pair. The slow motion of the heavy quarks is embedded in the potential which is an effective potential known as the Born-Oppenheimer potential \cite{Maiani:2022qze}.

The configuration of the tetraquark state in the dynamical diquark model originates from hybrid heavy-quark mesons; spatially extended colored field connecting a heavy color-\textbf{3} $\overline{\delta}$ antidiquark source  and a heavy color-$\overline{\textbf{3}}$  $\delta$ diquark source \cite{Lebed:2017min}. These sources are expected to be compact due to the presence of heavy quark but still can have a finite spatial extent which is $\lesssim 0.5 ~ \text{fm}$ for charm \cite{Brodsky:2014xia}. Although the $cc \bar c \bar c$ tetraquark state is not studied in this present paper, we observed that the distance between $cc$ diquark and $\bar c \bar c$ antidiquark is $\langle r \rangle \simeq 0.3 ~ \text{fm}$. For $cc$ and $\bar b \bar b$, this distance turns out to be as $\langle r \rangle \simeq 0.2 ~ \text{fm}$. This tetraquark system is compact and has slow motion as a whole due to the heavy quark content. Since we assume that diquarks are pointlike particles, we are not able to obtain the size of the $cc$, $cb$, and $bb$ diquarks. 

In the dynamical quark model, the definition of mass eigenstate is elevated with the gluon field connecting to diquarks. %In the diquark model of this work \cite{Giron:2019bcs}, exotic states are constructed from heavy light diquarks $\delta$, $\overline{\delta}$ which are formed in the attractive channels $ 3 \otimes 3 \to \overline{3}$ and $\overline{3} \otimes \overline{3} \to 3$ between the color-triplet quarks. 
Corresponding mass $m_{\delta}$ of the diquark is not a fundamental Lagrangian parameter so that $m_{\delta}$ and $\delta-\overline{\delta}$ potential are taken into account phenomenologically. The confinement limits the separation of diquark pairs and the specific stationary states of the full system are maintained by the quantized modes of the gluon field stretching between the two heavy, approximately stationary heavy diquark sources, $\delta$ and $\overline{\delta}$. In the case of fully-heavy tetraquark, this approach makes Born-Oppenheimer approximation usable in the same manner as is done for simulations of heavy-quark hybrids on the lattice studies, see for example Refs. \cite{Juge:1997nc,Juge:1999ie,Capitani:2018rox}. The gluon field corresponds to ``electrons" and slow heavy diquarks correspond to the ``nuclei". This is the essence of the dynamical diquark model and was the logic for the study of $c \bar c c \bar c$ in the dynamical diquark model with Born-Oppenheimer potentials which gave reliable results \cite{Giron:2020wpx}. In this present paper, we use the same logic and apply Born-Oppenheimer approximation to the tetraquarks made up by both $b$ and $c$ quarks.

\section{\label{sec:level5}Concluding Remarks and Final Notes}
The diquark models in the literature approximate the quasiparticles $\delta$ and $\bar{\delta}$ to be pointlike particles. However, it is important to point out that diquarks and antidiquarks are not pointlike objects. Diquarks are expected to have spatial extents which are comparable to the mesons with the same valence quark flavor content. Nonetheless, it is shown in Ref. \cite{Giron:2019cfc} that for $c\bar{c}q\bar{q}$ states the diquark size has a mild effect on the spectrum. 

The search for all-heavy tetraquark states is an ongoing topic. If a resonant state in the channel of two heavy quarkonia can be observed in the experiments, the tetraquark nature is favored for the inner structure of this resonant state. An indication for the di-$\Upsilon(1S)$ $b\bar{b}b\bar{b}$ state was reported in \cite{CMS:2016liw}. Two fully-heavy tetraquark candidates reported by LHCb in the di-$J/\psi$ ($c\bar{c}c\bar{c}$) \cite{Aaij:2020fnh}. For the time being, no structures have yet been reported in other all-heavy channels, such as $b\bar{b}c\bar{c}$, $b\bar{c}b\bar{c}$,  etc. 

In this work, we have investigated mass spectra for the $cc\bar{b}\bar{b}$, $bc\bar{c}\bar{c}$, and $bc\bar{b}\bar{b}$ tetraquark states in the dynamical diquark model. This model is the application to the dynamical diquark picture that describes a multiquark exotic state as a system of a compact diquark $\delta$ and antidiquark $\bar{\delta}$ for a tetraquark state. We also calculated root mean square radii $\sqrt{\langle r^2 \rangle} $ values. We observed that $cc\bar{b}\bar{b}$, $bc\bar{c}\bar{c}$, and $bc\bar{b}\bar{b}$
tetraquark states are compact.

In the framework of dynamical diquark model, $cc\bar{b}\bar{b}$ mass value is found to be below corresponding to the two-meson thresholds. This conclusion is supported by Ref. \cite{Wang:2021taf} for some interpolating currents and for $J^{PC}=0^{++}$ of Ref. \cite{Bedolla:2019zwg}. Ref. \cite{Anwar:2017toa} gave an upper limit for mass of the $cc\bar{b}\bar{b}$ tetraquark as 12.58 GeV which is below the $2B_c$ threshold indicating that $cc\bar{b}\bar{b}$ tetraquark is stable against strong decays. It was mentioned in Ref. \cite{Lebed:2022gfb} that $cc\bar{b}\bar{b}$ observation via $J/\psi-\Upsilon$ should produce many resonances, by evading the identical-fermion constraint. For $bc\bar{c}\bar{c}$ and $bc\bar{b}\bar{b}$ tetraquark states, all the predicted mass values are above the corresponding two-meson thresholds. These states should be broad and hard to be observed in the experiments. Similar conclusions were achieved in Refs. \cite{Deng:2020iqw,Faustov:2020qfm,Liu:2019zuc}. However, the lowest $1^+$ $bc\bar{b}\bar{b}$ tetraquark state was found to be good candidate of being stable \cite{Wu:2016vtq}.   

Regarding $cc\bar{b}\bar{b}$ state, it should be noted that the obtained result for the $cc$ diquark mass is slightly larger and $bb$ diquark mass is slightly smaller compared to the results of other approaches. The lowest state remains stable unless the diquark masses turn out to be approximately 30\% larger than those given in Eq. \ref{masssr}. Even some excited states of $B_c$ escape from the experiments, if some states in the $B_c$  meson spectra can be observed in future, our results may help for searching corresponding two-meson thresholds with respect to $cc\bar{b}\bar{b}$ system. 

Existing charmonium, bottomonium and $B_c$ states may pave the way for investigating resonance structures above the $\eta_c B_c$ threshold. Many theoretical studies have been devoted to fully-heavy tetraquark systems. The interpretations on the natures of these systems are not completely consistent in the different theoretical models. Therefore, more data are needed to enlighten the inner structures of these states. This will also provide a deeper understanding of the interactions of in the multiquark systems. We hope that the
decay channels which are discussed in this work may be helpful for the experimental search.

\textit{Note added-} Very recently, LHCb collaboration reported about observation of two open-flavor tetraquark states, $T_{c\bar{s}0}^a (2900)^0$ and $T_{c\bar{s}0}^a (2900)^{++}$ in the decays $B^0 \to D^0 D_s^+ \pi^-$ and $B^+ \to D^- D_s^+ \pi^+$ \cite{results}. The quark contents are  $c\bar s u \bar d$ for $T_{c\bar{s}0}^a (2900)^0$ and $c\bar s \bar u d$ for $T_{c\bar{s}0}^a (2900)^{++}$. These are the tetraquarks made up by four different valence quarks with only one heavy quark source. This makes these states more interesting since they may serve a laboratory for diquark-antidiquark and hadronic molecule pictures. Furthermore, the validity of BO approximation in the diquark schemes may be studied for these states since there exist only one heavy quark as a valence quark. The constituent quark mass of charm quark is around 1.5 GeV, strange quark mass is around 400 MeV, and up and down quark masses are around 300 MeV. The separation of masses $M_Q>> M_q$ makes BO potential to be used. In addition to this, relativistic effects should be taken into account due to the existence of light quarks. The compactness and effects of the diquark separation should be important for these newly observed particles. A detailed elaboration should be made on these states in terms of the diquark models mentioned above, which is in our agenda.

\begin{acknowledgments}
The author would like to thank to R. F. Lebed and J. F. Giron for their helpful discussions on the topic. This work is partially supported by The Scientific and Technological Research Council of Turkey (TUBITAK) in the framework of BIDEB-2219 International Postdoctoral Research Fellowship Program.
\end{acknowledgments}

\bibliography{ccbb}

\end{document}